\documentclass[sigconf,authorversion,nonacm]{acmart}

\usepackage{graphicx} 
\usepackage{xcolor}
\usepackage{hyperref}
\usepackage[capitalize]{cleveref}
\usepackage{booktabs}

\newcommand{\fvec}[1]{\pmb{#1}}

\graphicspath{{./resources/}}

\begin{document}

\title{XAI-based Comparison of Input Representations for Audio Event Classification}
\author{Annika Frommholz}
\email{annika.frommholz@hhi.fraunhofer.de}
\affiliation{%
  \institution{Fraunhofer Heinrich-Hertz Institute}
  \streetaddress{Einsteinufer 37}
  \city{Berlin}
  \country{Germany}
}

\author{Fabian Seipel}
\email{f.seipel@campus.tu-berlin.de}
\affiliation{%
  \institution{Technische Universit{\"a}t Berlin} 
  \streetaddress{}
  \city{Berlin}
  \country{Germany}
}

\author{Sebastian Lapuschkin}
\orcid{1234-5678-9012}
\email{sebastian.lapuschkin@hhi.fraunhofer.de}
\affiliation{%
  \institution{Fraunhofer Heinrich-Hertz Institute}
  \streetaddress{Einsteinufer 37}
  \city{Berlin}
  \country{Germany}
}

\author{Wojciech Samek}
\authornote{Corresponding author}
\email{wojciech.samek@hhi.fraunhofer.de}
\affiliation{%
  \institution{Fraunhofer Heinrich-Hertz Institute,\\ Technische Universit{\"a}t Berlin \& \\ BIFOLD - Berlin Institute for the Foundations of Learning and Data}
  \streetaddress{Einsteinufer 37}
  \city{Berlin}
  \country{Germany}
}

\author{Johanna Vielhaben}
\authornotemark[1]
\email{johanna.vielhaben@hhi.fraunhofer.de}
\affiliation{%
  \institution{Fraunhofer Heinrich-Hertz Institute}
  \streetaddress{Einsteinufer 37}
  \city{Berlin}
  \country{Germany}
}

\date{April 2023}

\begin{abstract}
    Deep neural networks are a promising tool for Audio Event Classification. In contrast to other data like natural images, there are many sensible and non-obvious representations for audio data, which could serve as input to these models. Due to their black-box nature, the effect of different input representations has so far mostly been investigated by measuring classification performance. In this work, we leverage eXplainable AI (XAI), to understand the underlying classification strategies of models trained on different input representations.  Specifically, we compare two model architectures with regard to relevant input features used for Audio Event Detection: one directly processes the signal as the raw waveform, and the other takes in its time-frequency spectrogram representation. 
    We show how relevance heatmaps obtained via "Siren"{Layer-wise Relevance Propagation} uncover representation-dependent decision strategies. With these insights, we can make a well-informed decision about the best input representation in terms of robustness and representativity and confirm that the model's classification strategies align with human requirements.  
\end{abstract}

\maketitle

\section{Introduction}
Audio Event Classification (AEC) is essential in applications such as audio scene recognition, robot navigation, or safety aids for the hearing-impaired \cite{vivek2020_hearingaid1, sun2021emergency, RAMIREZ2022_hearingaud2}. It involves recognizing and classifying specific sound events or patterns in an audio signal, such as everyday sounds or urban soundscapes. Due to the complex unstructured sounds that overlap at varying loudness levels and in many cases have poor recording quality, AEC is often more challenging than speech classification.

Machine Learning algorithms, particularly deep learning techniques, have shown promising results in AEC. Here, waveforms and spectrograms models are two common input representations to the model. Waveform-based models directly process the raw audio signal as a time series. In contrast, for spectrogram-based models, the audio signal is transformed to a two-dimensional image-like format, capturing the audio signal's frequency and time information. While both model types have been applied successfully for AEC tasks, little attention has been given to the differences the underlying classification strategies. 
With this work, we step into this gap and apply Layer-wise Relevance Propagation (LRP) \cite{lapuschkin2015_lrp}, a popular XAI technique \cite{samek_XAI_2021} to reveal the inner workings of the models and compare their classification strategies. With LRP, we can quantify the relevance of each input feature, here single time-points or the time-frequency components of a spectrogram, by propagating the model output back to the input. In particular, we leverage the recently proposed DFT-LRP, which first injects a virtual Discrete Fourier Transform (DFT) layer into the model and then applies LRP in order to gain interpretable access in a human-understandable latent representation within the model. DFT-LRP \cite{vielhaben2023explainable} enables the comparison of classification strategies of models trained on different input representations by transforming relevance heatmaps from time to time-frequency or frequency domain.

In summary, we investigate the classification strategies of two convolutional neural networks, one of which operates on the raw waveforms in time domain, and another one operating on time-frequency spectrogram representations for a popular audio event classification task. To this end, we compute relevance heatmaps in time-frequency domain for both model types using LRP and DFT-LRP, which quantify the importance of each time-frequency component toward the model output probability for a given class. These insights allow us to choose the most suitable input representation for AEC models not only based on classification performance but also considering the underlying model processes. Further, the XAI analysis reveals whether the model reasoning aligns with human requirements or if it has learned to base its decision on spurious correlations in the data.

\section{Related Work}
\label{sec:related_papers}

\paragraph{Audio Event Classification with different Input Representations}

In recent years deep neural networks have outperformed traditional classification models like Hidden Markov Models or Support Vector Machines on sound classification tasks \cite{kons2013_gmm_svm}. Most notably, multiple neural network architectures that use different audio data representations as model input have been developed. Starting with handcrafted acoustic features \cite{salamon2015_audio_feature_model}, to images of (Mel-) spectrograms \cite{piczak2015logmelCNN, Hershey2017CNN, tokozume2017CNN} or one-dimensional audio signals \cite{lee2017waveCNN} \cite{sang2018CRNN}, there are now also hybrid approaches \cite{wang2021multiformat} for audio classification. \\
Recently, \cite{tsalera2021comparison} trained spectrogram and waveform CNNs -  GoogLeNet, SqueezeNet, ShuffleNet, VGGish, and YAMNet, - on three sound datasets, among which the two waveform CNNs performed best (average 96.4\% accuracy) on all datasets.
A more detailed investigation of 1D-CNNs was carried out by \cite{abdoli2019_1dcnn} on the \textit{UrbanSound8k} dataset. Different window functions of convolutional filters and input lengths were tested for their effect on classification accuracy. The best result, also compared to 2D spectrogram CNNs, was achieved by initializing the first convolutional layer as a gamma tone filter bank and using a rectangular window with a length of 16000 time steps.

\paragraph{Explainable AI for Time Series Data}
\label{subsec:2_b_rel_papers_XAI}

An recent overview of XAI methods suited explicitly for time series data is given by \cite{rojat2021_overviewXAI}. They present different types of explanations and their influence on the stakeholder's confidence in AI systems.
In \cite{becker2018AudioMNIST}, LRP is used to explain two neural networks, one trained on spectrograms (AlexNet \cite{alexnet}) and the other on one-dimensional audio signals (AudioNet \cite{becker2018AudioMNIST}) of two simple speech classification tasks, and relevant input features show, that AlexNet uses different areas of the spectrogram for classifying the gender of the speaker. Most notably, the interpretability and comprehensibility of relevance heatmaps for waveform signals in the time domain only was worse than those for spectrogram representations.
Further, the work of \cite{colussi2021LRP} applies LRP to ambient noise recognition and compares two spectrogram representations regarding classification accuracy and class-specific relevant frequency ranges. On the \textit{UrbanSound8k} dataset, two adapted AlexNets are trained, with data being fed into the network once as a mel-spectrogram and once as a constant-Q-spectrogram. Using LRP relevance maps, ten noise classes are interpreted concerning class-related important frequencies.
In another line of research, \cite{zinemanas2021XAI} presents an intrinsically interpretable model architecture for speech, music, and audio event classification. This hybrid of autoencoder and classifier uses the embedding layer of the autoencoder and compares it to a prototype embedding per class. 

\section{Explainable AI for Audio Classification}
\label{sec:xai_audio}
We structure this methods section into three parts. First, we introduce the input representations that the two AEC models studied later will be based on. Second, we  describe LRP, which we use to quantify the importance of each input feature toward the classifier output for a given sample. Third, we lay out how LRP can be applied to layers implementing a DFT (and inverse DFT respectively) via the DFT-LRP approach, which provides a unified point-of-access in terms of the aforementioned input representations and enables a comparison between classification strategies of models trained on them.

\subsection{Input representations for Audio Classifiers}
\begin{figure*}[ht]
    \center
    \includegraphics[width=0.9\textwidth]{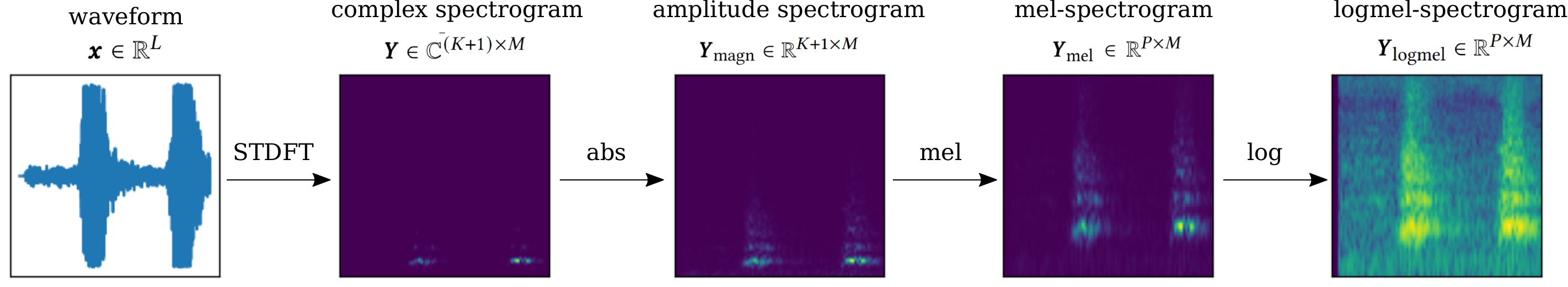}
    \caption{Stepwise transformation between raw waveform $x$ of the audio signal and logmel-spectrogramm $Y_{\text{logmel}}$.}
    \label{fig:signal_representations}
\end{figure*}

\paragraph{Waveform}
In time domain, an audio signal is represented by the waveform $\fvec{x} \in \mathbb{R}^{L}$, which contains its temporal amplitude values. The discrete time steps between the signal values depend on the sampling frequency $f_S$, and the signal duration is $\frac{L}{f_S}$. An example for a waveform representation is given in \Cref{fig:signal_representations} (leftmost panel).

\paragraph{Spectrogram}
\label{subsec:sinal_spectrogram}
From the raw waveform $\fvec{x}$, we can extract information about the frequency content varying over time by applying the \textit{Short Time Discrete Fourier Transform (STDFT)},
\begin{equation}
    STDFT(\fvec{x}) = Y_{k,m} = \sum_{n=0}^{N-1}x_{n+mH} \cdot w_{n} \cdot e^{-\frac{i \pi kn}{N}}
    \label{eq:stdft}
\end{equation}
Here, a Discrete Fourier Transform is calculated for potentially overlapping windowed parts of the signal, depending on length $M$ and hop size $H$ of the window function $w$.
This yields the signal representation in time-frequency domain, i.e. the \textit{Spectrogram} $\fvec{Y} \in \mathbb{C}^{(K+1) \times M}$, which contains the complex-valued time-frequency components in $K+1$ frequency and $M$ time bins with $K=\frac{N}{2}$ and $M = \frac{L-N}{H}$.
For most classification applications, we can disregard the phase information and only consider the amplitude $\fvec{Y}_{\text{magn}} \in \mathbb{R}^{K+1 \times M}$  of the complex spectrogram. 

Now, we follow  \cite{vggish_yamnet_precessor} and bring the spectrogram to the mel-scale, which is a "melodic" scaling of the frequencies that accounts for the psychoacoustic phenomenon that humans do not perceive the same ratio between two frequencies in the same way for low frequency and high frequency ranges \cite{weinzierl2009_handbuchAT}. For example, a frequency doubling from 250\;Hz to 500\;Hz is perceived as a doubling of pitch, whereas a perceived pitch doubling of 1300\;Hz corresponds to an actual frequency of 8000\;Hz. In the mel-scale, the perceived pitch differences are equally distanced. Compared to linear frequency scaling, the mel-scaling widens low frequencies and compresses high frequencies to mimic human perception of frequency ratios. A mel-scaling $\text{mel}(\cdot)$ of frequencies can be obtained by multiplying the magnitude spectrum $\fvec{Y}_{\text{magn}}$ with a triangular filterbank $\fvec{T} \in \mathbb{R}^{K+1 \times P}$
with $P$ triangular filters decrease in height and increase in width with higher frequencies. The linear middle frequencies of the triangular filters are equally distanced in the mel-scale. 
The resulting mel-spectrogram $\fvec{Y}_{\text{mel}} \in \mathbb{R}^{P \times M}$ contains the time-frequency information of the time series in $P$ mel-scaled frequency bins and $M$ time bins. Finally, to account for the logarithmic nature of signal loudness levels, we take the logarithm of the mel-spectrogram amplitude values to obtain a logmel-spectrogram $\fvec{Y}_{\text{logmel}} \in \mathbb{R}^{P \times M}$, again following \cite{vggish_yamnet_precessor}.

The above steps that connect waveform $\fvec{x}$ and the logmel-spectrogram $\fvec{Y}_{\text{logmel}}$ are illustrated in \Cref{fig:signal_representations}.

\subsection{Layer-wise Relevance Propagation}
\label{subsec:lrp}

\textit{Layer-wise Relevance Propagation (LRP)} \cite{lapuschkin2015_lrp} is a \textit{XAI} method that uses modified backpropagation iteratively through all layers of the network to distribute relevance scores to the input features based on their importance to the model's prediction at the final output layer.
In general, the relevance score $R_j$ of a neuron in an upper layer $j$ is fully distributed onto all neurons from a lower layer $i$,
\begin{equation}
\label{eq:lrp}
    R_{i} = \sum_{j} R_{i \leftarrow j} = \sum_{j} z_{ij} \frac{R_j}{\sum_{j} z_{ij}} 
\end{equation}
 where $z_{ij}$ denotes a pre-activation resulting from a lower layer activation $a_i$ in interaction with a model parameter $w_ij$.
Different choices for $z_{ij}$ lead to different LRP rules \cite{montavon2019_lrp_overview}  that can be combined to account for the specific model architecture \cite{KohIJCNN20}.
A simple choice is $z_{ij}=w_{ij}a_i$ and corresponds to the LRP-${\varepsilon}$ rule, which also adds a small value $\varepsilon$ in the denominator of \Cref{eq:lrp} for numerical stability. The LRP-$z^+$ rule \cite{Montavon2017_zrule} only considers positive parts of the pre-activations $z_{ij}$ in Equation \Cref{eq:lrp} turns into $z_{ij}=z_{ij}^+=(a_iw_{ij})^{+}$ yielding only positive relevances $R_i$. The combination of these two rules is defined as the LRP-$\varepsilon^+$ propagation rule where LRP-$z^+$ is applied to convolutional layers and LRP-$\varepsilon$ to dense layers, and is known to improve qualitative as well as quantifyable attributes of relevance maps \cite{KohIJCNN20}. Insignificant and contradictory relevances get absorbed, leading to sparser relevance maps where mostly strong relevances contribute.
In general, relevances can be positive or negative. Positive values in the activation map highlight features which have a relevant impact on the classification decision in favor of a given class. Negative values mean that a feature is contradictory to the model's prediction, i.e. speak against a specific class.   

\subsection{DFT-LRP propagates relevances to different input representations}
\label{subsec:dft-lrp}

\begin{figure}[htbp]
    \centerline{\includegraphics[width=0.5\textwidth]{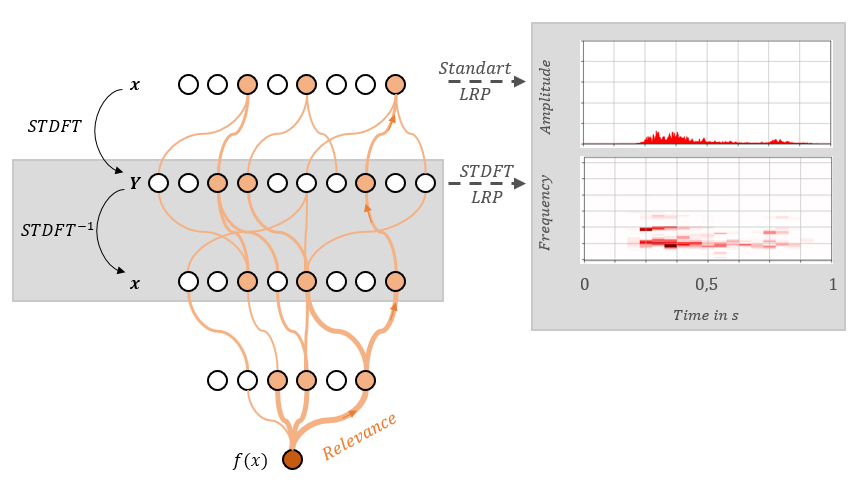}}
    \caption{DFT-LRP for propagating relevance from time to time-frequency domain.}
    \label{fig:dft_lrp}
\end{figure}

For audio samples in the form of waveforms in time domain, relevance heatmaps are hard to interpret, and inspecting the \mbox{(time-)frequency} domain is often more comprehensive for users. To transform the relevance information in time domain into a time-frequency representation, we leverage the \textit{DFT-LRP} method from \cite{vielhaben2023explainable}. The basic idea is that since the DFT and STDFT are linear transformations, LRP can be applied. 
Before inference (predicting the class for a new sample by forward passing through the model), two layers are added to the network after the input layer. An identity loop consisting of a \textit{Discrete Fourier Transform} $\text{DFT}(\cdot)$ and \textit{Inverse Discrete Fourier Transform} $\text{IDFT}(\cdot)$, acting as a virtual inspection layer, ensure that the forward pass of the input data also runs through the frequency domain so $\fvec{x}' = \text{IDFT}(\fvec{Y}) = \text{IDFT}(\text{DFT}(\fvec{x}))$. As illustrated in \Cref{fig:dft_lrp}, when backpropagating the relevance through all layers as described above, they also pass through the $IDFT$ layer and can be extracted at this point for relevances in the time-frequency domain. The LRP-rule for relevance propagation through the STDFT in \Cref{eq:stdft} is,
\begin{equation}
    \label{eq:stdft-lrp}
    R_{k,m} = Y_{magn,k,m} \sum_n \cos(\frac{2 \pi k n}{N} - \varphi_{m,k}) \cdot  w_n^{-1} \frac{R_n}{x_n}
    \, ,
\end{equation}
where $R_n$ is the relevance on the signal in time domain $x_n$, and $\varphi_{m,k}$ is the phase information of $Y_{m,k}$ \cite{vielhaben2023explainable}.

\section{Results}

First, we qualitatively compare exemplary heatmaps between the two models. Second, we compare classification strategies by correlating heatmaps. Third, we build on the visual impression of heatmaps and test robustness towards audio augmentations as high- and low-pass filters.

\subsection{Dataset and models}
We introduce the dataset used in our experiments, and the two model architectures, that process raw waveforms and mel-spectrograms, respectively, and that we will investigate in the following sections. 

\paragraph{UrbanSound8k Dataset} We base all of our experiments on the \textit{UrbanSound8k} audio event classification dataset \cite{salamon2014_urbansound8k}. It contains \textit{wav}-files ($\le 4s$) for ten urban sound classes, namely "Air Conditioner", "Car Horn", "Children Playing", "Dog Bark", "Drilling", "Engine Idling", "Gun Shot", "Jackhammer", "Siren", and "Street Music". The dataset consists of 8732 audio samples of varying duration which sum up to 7.3 hours of material. 
We follow \cite{abdoli2019_1dcnn} and apply the following pre-processing steps: First, the audio files of different lengths, sample rates, and channels are unified to 1s-long patches with a sampling rate of $f_S=16 kHz$, overlapping by 50\% and converted to mono. In this way, we create $38000$ samples $\fvec{x} \in \mathbb{R}^{16000}$. 
The UrbanSound8k dataset is provided with a signal range $[-1,1]$, but to account for the different types of sounds, we normalize each sample by its root mean square. In this way, quiet continuous noise-like sounds have a different dynamic range than short, loud events. 
Further, during model training, we use the following additional audio augmentation to promote robustness and increase generalization performance:
\begin{itemize}
    \item Gain between $-12dB$ and $-1dB$.
    \item Noise with signal-to-noise-ratio between $10^{-4}$ and $10^{-1}$.
    \item Delay between $1ms$ and $300ms$.
    \item Bandpass filter with cutoff frequencies in the intervals $f_{C,LP}=[1400Hz;4000Hz]$ and $f_{C,HP}=[500Hz;1200Hz]$ for a low pass (LP) and high pass (HP) filter, respectively.
\end{itemize}

\paragraph{1DCNN} This convolutional network architecture introduced by \cite{abdoli2019_1dcnn} processes audio sample as one-dimensional raw waveforms in time domain. It includes four convolutional layers and contains $2.5\times 10^6$ parameters in total. We train 1DCNN on the UrbanSound8k audio event classification task and achieve a training set accuracy of $75.22 \pm 0.44 \%$, and test set accuracy of $54.94 \pm 5.19 \%$ over a 10-fold cross-validation.

\paragraph{YAMNet (Yet Another Merging NETwork)} This is a convolutional neural network, that processes audio samples as two-dimensional logmel-spectrograms in the time-frequency domain. It is built on the architecture of
Mobilenet V1 \cite{howard2017_mobilenet}. This applies depthwise-separable convolutions, that allow for an efficient calculation of 2D convolutions. YAMNet has 13 convolutional layers and $32.1\times 10^6$ parameters in total.
Before training YAMNet on the Urbansound8k audio event classification task, we need to transform the signal from waveforms to logmel-spectrogams.  For the STDFT, see \Cref{eq:stdft}, we choose a rectangular window with a length of 800 time steps and a hop size of $H=800$. After applying the STDFT and taking the magnitude, we get a spectrogram $\fvec{Y}_{\text{magn}} \in \mathbb{R}^{8001 \times 20}$. Next, we apply the mel transform in \Cref{eq:mel_scale}  with $P=64$ triangular filter. After taking the logarithm, the logmel-spectrogram $\fvec{Y}_{\text{logmel}} \in \mathbb{R}^{64 \times 20}$ is used as model input for the two-dimensional YAMNet. 
In a 10-fold cross-validation, the model achieves a training accuracy of $75.22 \pm 0.44 \%$, and a test accuracy of $53.52 \pm 8.41 \%$.

\subsection{Qualitative analysis of 1DCNN and YAMNet classification strategies}
XAI allows us to compare classification strategies between different models. Here, we start with an exemplary qualitative analysis of 1DCNN and YAMNet  classifications based on LRP relevance heatmaps. In particular, for comparability of classification strategies, we leverage DFT-LRP to obtain relevance heatmaps for 1DCNN and YAMNet  which process input in different domains, i.e. time and time-frequency domain, in the same domain, i.e. time-frequency domain.

We compute LRP relevance heatmaps for true class logits of 1DCNN and YAMNet for the test set samples of the UrbanSound8K dataset. Here, we employ the LRP-$\varepsilon^+$ rule, see section \Cref{subsec:lrp}, and use the PyTorch LRP implementation \textit{zennit} \cite{anders2021_zennit}. 
To be able to compare classification strategies between the models, we require relevances from both models in the same representation and choose the mel-spectrogram in time-frequency domain. Since the YAMNet already receives the input data as a logmel spectrogram $\fvec{Y}_{\text{logmel}} \in \mathbb{R}^{64 \times 20}$ no further processing of the heatmaps is required. In contrast, the original input to the 1DCNN is the raw waveform. Thus, we apply DFT-LRP, to transform 1DCNN heatmaps from time to time-frequency representation. 
Then, we convert them to the to mel frequency scale with 64 bands, finally yielding relevance maps $\fvec{R}_{\text{mel}} \in \mathbb{R}^{64 \times 20}$. 

For a first qualitative comparison of classification strategies, we evaluate the middle frequency of the \textit{most relevant frequency bin} $f_{\text{rel}}$, which gives information about the per-class frequency focus. To compute $f_{\text{rel}}$, we average the most relevant frequency bin of relevance heatmaps $\fvec{R}^n_{\text{mel}}$ over all samples $n$ of a a class $C$ with $N_C$ samples $f_{\text{rel},C}$,
\begin{equation}
    f_{\text{rel},C} = \text{lin}_m\left[ \frac{1}{N_C} \sum_{n=0}^{N_C-1} argmax \left( \sum_{m=0}^{M-1} R^n_{mel,p,m} \right) \right] \, ,
    \label{eq:f_rel}
\end{equation}
where $\text{lin}_m$ is the mel bin middle frequency in Hz.

\begin{table*}
    \centering
    \caption{Qualitative analysis of LRP relevance heatmaps for 1DCNN and YAMNet.}
    \begin{tabular}{clc|cr|cr}
        \multicolumn{3}{c}{}& \multicolumn{2}{c}{\textbf{1DCNN}} & \multicolumn{2}{c}{\textbf{YAMNet}}  \\
        \toprule
         \multicolumn{2}{l}{Class} & Count   & accuracy &  $f_{\text{rel}}$ &         accuracy & $f_{\text{rel}}$ \\
        \midrule
            0 &  "Air Conditioner" &    600   &   89.17\% &   201 Hz    & 40.17 \% &  273 Hz    \\
            1 &  "Car Horn" &    124   &   87.10\% &   791 Hz   & 85.48 \% &  201 Hz    \\
            2 &   "Children Playing" &    596   &   65.44\% &   528 Hz   & 64.43 \% &  482 Hz  \\\
            3 &  "Dog Bark" &    457   &   61.71\% &   680 Hz    & 68.05 \% &  393 Hz  \\\
            4 &  "Drilling" &    542   &   52.40\% &  1414 Hz  & 55.35 \% &  528 Hz  \\
            5 &  "Engine Idling" &    550   &   56.55\% &    71 Hz   & 70.18 \% &  236 Hz   \\
            6 &  "Gun Shot" &     80   &   77.50\% &   393 Hz    & 86.25 \% &  102 Hz   \\
            7 &  "Jackhammer" &    516   &   85.08\% &   482 Hz  & 60.85 \% &  393 Hz    \\
            8 &  "Siren" &    476   &   37.61\% &   577 Hz    & 46.01 \% &  482 Hz   \\
            9 &  "Street Music" &    600   &   41.17\% &   352 Hz    & 83.83 \% &  577 Hz   \\
        \bottomrule
    \end{tabular}
    \label{tab:model_results}
\end{table*}

\begin{figure*}
    \centerline{\includegraphics[width = 0.8\textwidth]{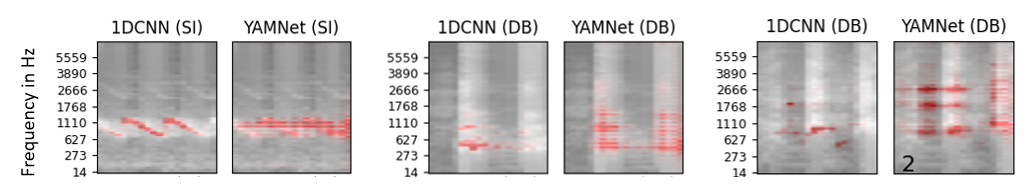}}
    \caption{Relevance maps for 1DCNN and YAMNet for three example audio samples. \textbf{Left:} "Siren" sample correctly classified by both models. \textbf{Middle:} "Dog Bark" sample correctly classified by both models. \textbf{Right:} "Dog Bark" sample correctly classified only by 1DCNN, the YAMNet predicted "Children Playing" (class 2).}
    \label{fig:qualitative_analysis}
\end{figure*}

We list the most relevant frequency bins for each class and both models in \Cref{tab:model_results} along with class-wise accuracy scores.
Classes, where the 1DCNN and the YAMNet assign most relevance to similar $f_{\text{rel}}$, are "Children Playing" and "Air Conditioner". 
For most other classes, YAMNet assigns the most relevance to lower frequencies than the 1DCNN. This is especially extreme in the case of "Drilling" and "Car Horn".\\ 
Further, we select three exemplary test samples and show relevance heatmaps for 1DCNN and YAMNet in \Cref{fig:qualitative_analysis} for an exemplary comparison of classification strategies.
The first heatmap pair in \Cref{fig:qualitative_analysis} shows a "Siren" sample correctly classified by both models. 
While both models successfully identified the relevant frequency range for this sound, the 1DCNN pays attention to the periodic change in frequency, and the YAMNet applies continuous relevance to the involved frequencies ignoring the temporal information. \\
Secondly, a correctly classified "Dog Bark" example shows the YAMNet's ability to localize multiple sound events in contrast to the 1DCNN focussing on the first bark. Nevertheless, the YAMNet focuses on the lowest narrow frequency bin despite the distinct events. Moreover, the 1DCNN assigns the majority of the relevance to the fundamental frequency whereas the YAMNet uses at least two overtones. \\
The last pair also shows a "Dog Bark", but the YAMNet misclassified it as "Children Playing" (2). Here, YAMNet bases its decisions primarily on pitch information. Visually the relevance map is very similar to the second example of a correctly classified "Dog Bark" with the main difference being the fundamental frequency. 

\subsection{Quantitative XAI-based comparison of 1DCNN and YAMNet classification strategies }
The choice of the input representation is an important question in deep learning based Audio Event Classification. Previous work only compares model trained in time or frequency domain based on classification accuracies \cite{tsalera2021comparison}. Again, we can leverage DFT-LRP relevance heatmaps in a uniform domain, i.e. time-frequency, to make a quantitative comparison of classification strategies between models that were trained in different domains, i.e. time and time-frequency domain.

We compute LRP relevance heatmaps for true class logits of 1DCNN and YAMNet like before.
To compare the classification strategies, first, we average the heatmaps over all test set samples within one class, in order to identify relevant frequency ranges and patterns used for different classes. Further, inspired by the Spectral Centroid \cite{peeters2004_spectralcentroid}, we compute a "Siren"{Relevance Centroid} $C_{R,m}$,  
which is the frequency-weighted mean of the relevances per time bin $m$,
\begin{equation}
    C_{R,m} = \frac{\sum_{p=0}^{P}f_{p} R_{p,m}}{\sum_{p=0}^{P}R_{p,m}} \, .
    \label{eq:relevance_centroid}
\end{equation}
Here, $f_{p}$ is the middle frequency of the mel bin $p$, and $R_{p,m}$ is the positive relevance value of this frequency-time bin. 
Second, we quantify similarities between heatmaps containing the relevance scores. We compute the cosine similarity $S_C= \fvec R^i\cdot \fvec R^j / (|\fvec R^i|\cdot |\fvec R^j|)$ between pairs of heatmaps $(\fvec R^i, \fvec R^j)$ after flattening them.
To account for the potential temporal shift between two sound events in the same class, we align heatmaps before calculating similarities. To this end, for each pair, both heatmaps are shifted temporally to the position of greatest cross-correlation between them. Lastly, we take the average over all heatmap pairs that are in the same class $\varnothing S_{C, within}$ and the average over all heatmap pairs of samples from different classes $\varnothing S_{C, between}$.

\begin{figure*}
    \centering
         \centering
         \includegraphics[width=\textwidth]{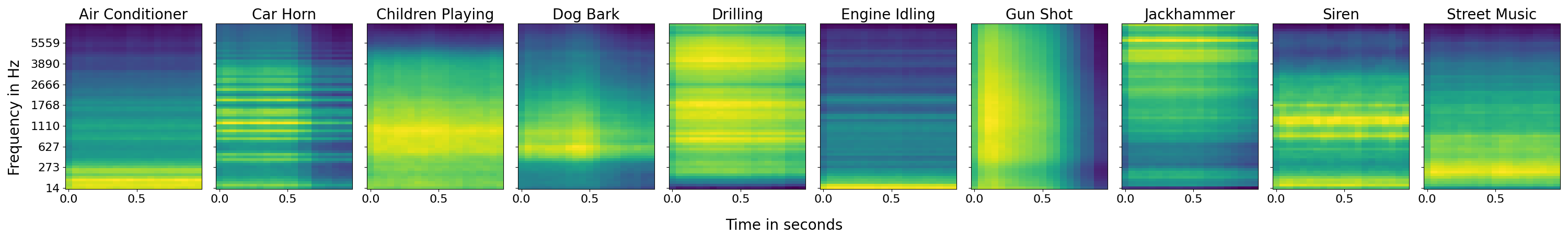}
         \includegraphics[width=\textwidth]{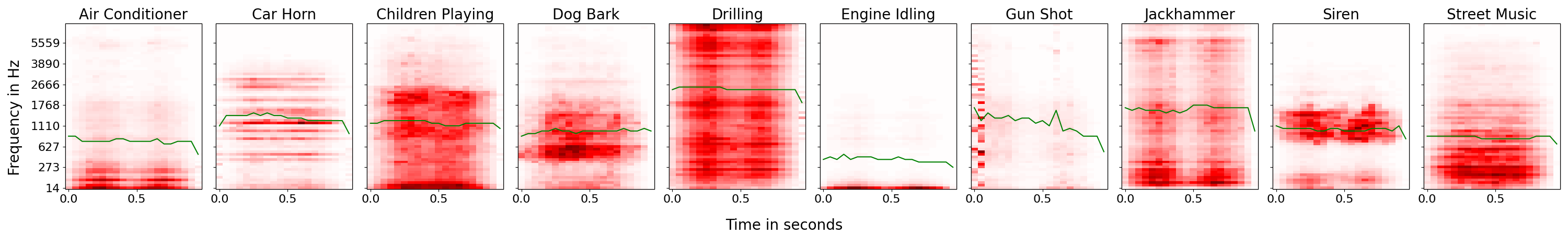}
         \includegraphics[width=\textwidth]{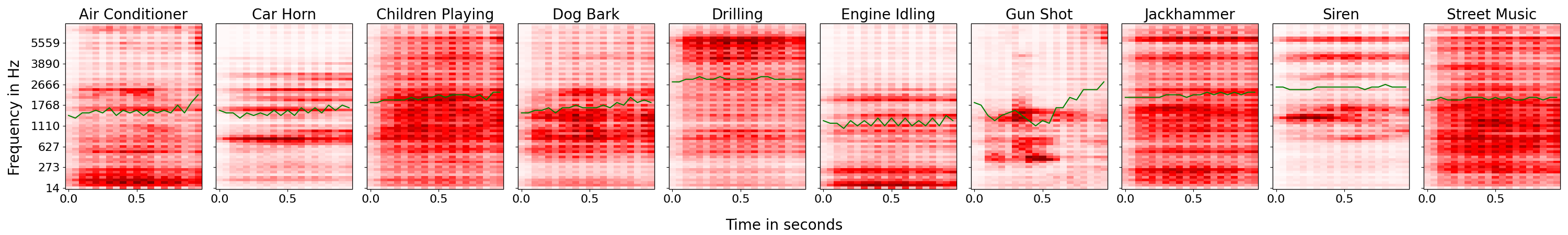}
    \caption{First row: Average spectrograms. Second row: per-class average test set  relevance heatmaps for 1DCNN. Third row: per-class average test set relevance heatmaps for YAMNet.}  
    \label{fig:avg_relevance_spectrograms}
\end{figure*}

We show the class-wise averaged heatmaps with Relevance Centroid $C_{R,m}$ in \Cref{fig:avg_relevance_spectrograms} and list the within and between class similarities of heatmaps in \Cref{tab:relevance_similarity_per_model}. The greater difference between the within-class similarity values $S_{C,\text{within}}$ and the between-class similarity values $S_{C,\text{between}}$ show that the 1DCNN bases the classification on input features that differentiate the classes. In contrast, the relevance maps of the YAMNet have high within-class and between-class similarity values. 
Thus, the 1DCNN has learned more separable class-specific characteristics of the ten sound classes, than YAMNet. 
The visual impression of the average heatmaps in \Cref{fig:avg_relevance_spectrograms} supports this finding. 
\begin{table}
    \centering
    \caption{Average "Siren"{within-class} and "Siren"{between-class} cosine similarity values (with standard deviation) for test set relevance maps of both model architectures.}
    \begin{tabular}{l|c|c}
        \toprule
        & \textbf{1DCNN} & \textbf{YAMNet} \\
        \midrule
        $\varnothing S_{C,\text{within}}$ & $0.207 \pm 0.057$ & $0.593 \pm 0.099$ \\ 
        $\varnothing S_{C,\text{between}}$ & $0.076 \pm 0.017$ & $0.584 \pm 0.062$ \\ 
        \bottomrule
    \end{tabular}
    \label{tab:relevance_similarity_per_model}
\end{table}

\subsection{Robustness against audio alteration}
Lastly, we test the robustness of the two models trained on different input representations against audio filtering and alteration that a human listener is indifferent to. 

To this end, we evaluate the test set accuracies for 1DCNN and YAMNet after applying a high pass filter with a cut-off frequency of 3000\;Hz and a low pass filter with a cut-off frequency of 3000\;Hz to the samples. Additionally, we choose siren as an example of a sound class with tonal character and measure the accuracy for this class after applying pitch shifting by seven 7 half-tones.
\begin{table}
    \centering
    \caption{Results for all pitch augmentations tested in the individual model analyses. The difference in test accuracy when applying a high pass ($f_C=300Hz$) or a high pass filter ($f_C=3000Hz$) to the complete test dataset and when applying a pitch-shift ($\pm$ 7 half-tones) to the "Siren" samples are shown.}
    \begin{tabular}{l|rr}
         & \multicolumn{2}{c}{\textbf{Accuracy Difference}}\\
         \midrule
        Augmentation & 1DCNN & YAMNet \\
        \midrule
        High Pass Filter                & -4.41\%   & -18.19\%\\
        Low Pass Filter                 & -3.85\%   & -8.48\% \\
        Pitch-Shift (on "Siren") & -9.67\%   & -25.00\%\\
        \bottomrule
    \end{tabular}
    \label{tab:results_augmentation}
\end{table}
We show the effects of these audio augmentations on the test set accuracies of 1DCNN and YAMNet in \Cref{tab:results_augmentation}.
For all three modifications, YAMNet suffers a greater drop in classification accuracy than 1DCNN. Thus, the 1DCNN classification performance and strategies are more robust against pitch-related augmentations than the YAMNet. In consequence, YAMNet, which uses the time-frequency representation of the input data relies more on pitch information for categorizing sounds than 1DCNN that processes the raw waveform. 
 
\section{Conclusion}
In this work, we leverage post-hoc XAI in the form of LRP to compare the classification strategies of convolutional neural networks trained on two different input representations of audio samples for a sound classification task. We find two major differences between their classification strategies.
First, The 1DCNN has learned more separable class-specific characteristics of the ten sounds, compared to the YAMNet, as revealed by the greater difference between the within-class similarity values $S_{C,\text{within}}$ and the between-class similarity values $S_{C,\text{between}}$.
Second, the effect of applying a low pass filter, a high pass filter,
and pitch-shifting shows that the classification performance of 1DCNN is more robust against pitch-related augmentations than for YAMNet, suggesting that the architecture that uses time-frequency representation of the input data relies on
more on pitch information for categorizing sounds.
These insights from the XAI-based model comparison not only help us to understand the underlying reasoning processes of the model, but they could also guide the design of future models specialized in audio applications, and aligning with human requirements.

\section{Acknowledgements}
This work was supported by the Federal Ministry of Education and Research (BMBF) as grants [SyReal (01IS21069B), BIFOLD (01IS18025A, 01IS18037I)]; the European Union’s Horizon 2020 research and innovation program under grants iToBoS (grant No. 965221) and TEMA (grant No. 101093003); the state of Berlin within the innovation support program ProFIT (IBB) as grant [BerDiBa (10174498)]; and the German Research Foundation [DFG KI-FOR 5363].

\bibliographystyle{ACM-Reference-Format}
\bibliography{main}

\end{document}